\documentclass[12pt,preprint]{aastex}
\usepackage{multirow}
\begin{document}
\title{Vortical Motions of Baryonic Gas in the Cosmic Web: Growth History and Scaling Relation}
\author{Weishan Zhu$^{1,2}$, Long-long Feng$^{1,3}$}
\altaffiltext{1}{Institute of Astronomy and Space Science, Sun Yat-Sen University
(SYSU), Guangzhou 510275, China}
\altaffiltext{2}{School of Astronomy and Space Science, Nanjing 
University, Nanjing, 210092, China}
\altaffiltext{3}{Purple Mountain Observatory, CAS, Nanjing, 210008, China}

\begin{abstract}
The vortical motions of the baryonic gas residing in large scale structures are investigated by cosmological hydrodynamic simulations. Proceeding in the formation of the cosmic web, the vortical motions of baryonic matter are pumped up by  baroclinity in two stages, i.e., the formation of sheets, and filaments. The mean curl velocity are about $< 1$, 1-10, 10-150, 5-50 km/s in voids, sheets, filaments and knots at $z=0$, respectively. The scaling of the vortical velocity of gas can be well described by the She-Leveque hierarchical turbulence model in the range of $l<0.65(1.50) h^{-1}$ Mpc in simulation of box size 25(100) $h^{-1}$ Mpc. The fractal Hausdorff dimension of vortical motions, $d$, revealed by velocity structure functions, is $\sim 2.1-2.3$($\sim 1.8-2.1$). It is slightly larger than the fractal dimension of mass distribution in filaments, $\textit{D}^f \sim 1.9-2.2$, and smaller than the fractal dimension of sheets, $\textit{D}^s \sim 2.4-2.7$. The vortical kinetic energy of baryonic gas is mainly transported by filaments. Both the scaling of mass distribution and vortical velocity increments show distinctive transition at the turning scale of $\sim 0.65(1.50) h^{-1}$ Mpc, which may be closely related to the characteristic radius of density filaments. 
\end{abstract}

\keywords{cosmology: theory - large-scale
structure of the universe - intergalactic medium - methods: numerical}

\section{Introduction}
In the concordance LCDM universe, the baryonic matter makes up 4.6\% of the critical cosmological density. About 10\% of the baryonic matter is in the form of collapsed objects such as stars and galaxies, and the rest is gas. Most of the baryonic gas is expected to reside in cosmic structures like sheets and filaments, and exists as inter-galactic medium(IGM) (see e.g., Fukugita \& Peebles 2004; Bregman 2007 and reference therein). The IGM feeds the growth of galaxies and receives feedback from star formation and active galactic nuclei. The properties of galaxies, including morphology and angular momentum and so on, would be closely related to the dynamics of ambient IGM. 

Analytical and simulation studies have investigated the dynamical evolution of IGM in the nonlinear regimes. In the mildly nonlinear regimes, the gas flow is approximately curl free and can be described by the stochastic-force driven Burgers equation. Moreover, it exhibits similar intermittent behaviors as turbulence (e.g. Shandarin \& Zeldovich 1989; Matarrese, \& Mohayaee 2002). Nevertheless, the curl free approximation becomes inapplicable in the highly nonlinear regime. Along with curved shocks emerging during the formation of cosmic structures, the baroclinity around shock fronts could effectively boost the generation of vorticity, which are then stretched and amplified by the shear tensor and divergence. The evolution of velocity power spectrum and vorticity in the IGM show characteristics of well developed turbulence under a couple of Mpc at low redshifts (Zhu et al. 2010,2013).  

The vorticity of baryonic matter might have non-negligible impact on the evolution of galaxies angular momentum and morphology. So far, numerous investigations have been made on the connections between large scale shear tensor, vorticity, spins of both dark matter halos and galaxies in various cosmic environments, i.e., voids, sheets, filaments and knots(clusters). The spins of dark matter halos were found to strongly aligned with the vorticity of ambient flow in simulations (Libeskind et al 2013; Wang et al. 2013; Laigle et al. 2015). Laigle et al. (2015) showed that the vorticity of the baryonic matter and the dark matter present a similar alignment to the filaments of the cosmic web in cosmological hydrodynamical simulations. Dubois et al. (2014) found that the spin orientation of their simulated galaxies is preferentially aligned with the vorticity of gas. The evolution of gas vorticity in the four category of cosmic structures, however, has not been studied in detail. 

On the other hand, the intermittency model of fully developed turbulence predicts that the fundamental scaling behavior of velocity fluctuations, i.e., velocity structure functions, is related to the most intensively hierarchical structures (She \& Leveque1994; Dubrulle 1994). As the vorticity and associated curl motions of the IGM show highly turbulent behavior, its velocity structure functions could be used to check the universality of the She-Leveque model (hereafter the SL model). Once checked out, the geometric information of the dominant structures of vorticity in the IGM derived from the SL model, in turn, can be confronted with the geometric properties of the cosmic web. Combing with the growth history, the scaling behavior of the curl velocity fluctuations could provide evident verification on the physical interpretation of the SL model, and insightful information on the transportation of vortical kinetic energy in the IGM during the formation of cosmic structures. 

In this paper, we study the vortical motions of the baryonic gas residing in large scale structures by cosmological hydrodynamic simulations. The growth history of vortical motions of gas in voids, sheets, filaments and knots are reported in \S2.  The scaling properties of the curl velocity increments, namely the velocity structure functions, and the connection to the cosmic web are investigated in \S3. Discussion and conclusion are given in \S4.   

\section{Vortical Motions of Gas in the Cosmic Web}
\subsection{Simulation Samples and the Cosmic Web}

The simulations in this paper were performed using the hybrid cosmological N-body/hydro-dynamics code WIGEON, which is based on the positivity-preserving Weighted Essentially Non-Oscillatory(WENO) finite differences scheme for the hydro-solver, and incorporated with the standard particle-mesh method for the gravity calculation (Feng et al. 2004; Zhu et al. 2013). In setting up the simulations, we adopt the WMAP5 normalization with the cosmological parameters $(\Omega_{m}, \Omega_{\Lambda},h,\sigma_{8} \Omega_{b},n_{s})=(0.274,0.726,0.705,0.812,0.0456,0.96)$(Komatsu et al. 2009). The simulations are evolved from redshift $z=99$ to present in periodic cubical boxes of side length 25 and $100\ h^{-1}$ Mpc,  hereafter referred to as L025 and L100 respectively.  With equal number of grid cells and dark matter particles of $1024^3$, the space and mass resolutions are determined to be $24.4 h^{-1}$ kpc/$1.3 \times 10^{6} \ M_{\odot}$, and $97.7 h^{-1}$kpc/$8.3\times 10^{7} \ M_{\odot}$ in L025 and L100 correspondingly. A uniform UV background is switched on at $z=11.0$ to mimic the re-ionization. The radiative cooling and heating processes are modeled with a primordial composition $(X=0.76,y=0.24)$. Star formation, AGN and their feedback are not implemented.

The matter distribution in simulated samples is classified into four categories of structures, i.e, voids, sheets, filaments and knots, identified with positiveness of eigenvalues of the potential tidal tensor (\textbf{d-web} hereafter), and the velocity shear tensor (\textbf{v-web}) following Hoffman et al. (2012). During the web identification processes, the baryonic density and velocity fields are smoothed with the Gaussian filter of radius of two grid cells, i.e., $48.8, 195.4 h^{-1}$ kpc in L025 and L100 respectively, and then are used to calculate the tidal tensor, and velocity shear tensor.

\begin{figure}[htbp]
\vspace{-0.5cm}
\hspace{-1.5cm}
\includegraphics[width=0.68\textwidth]{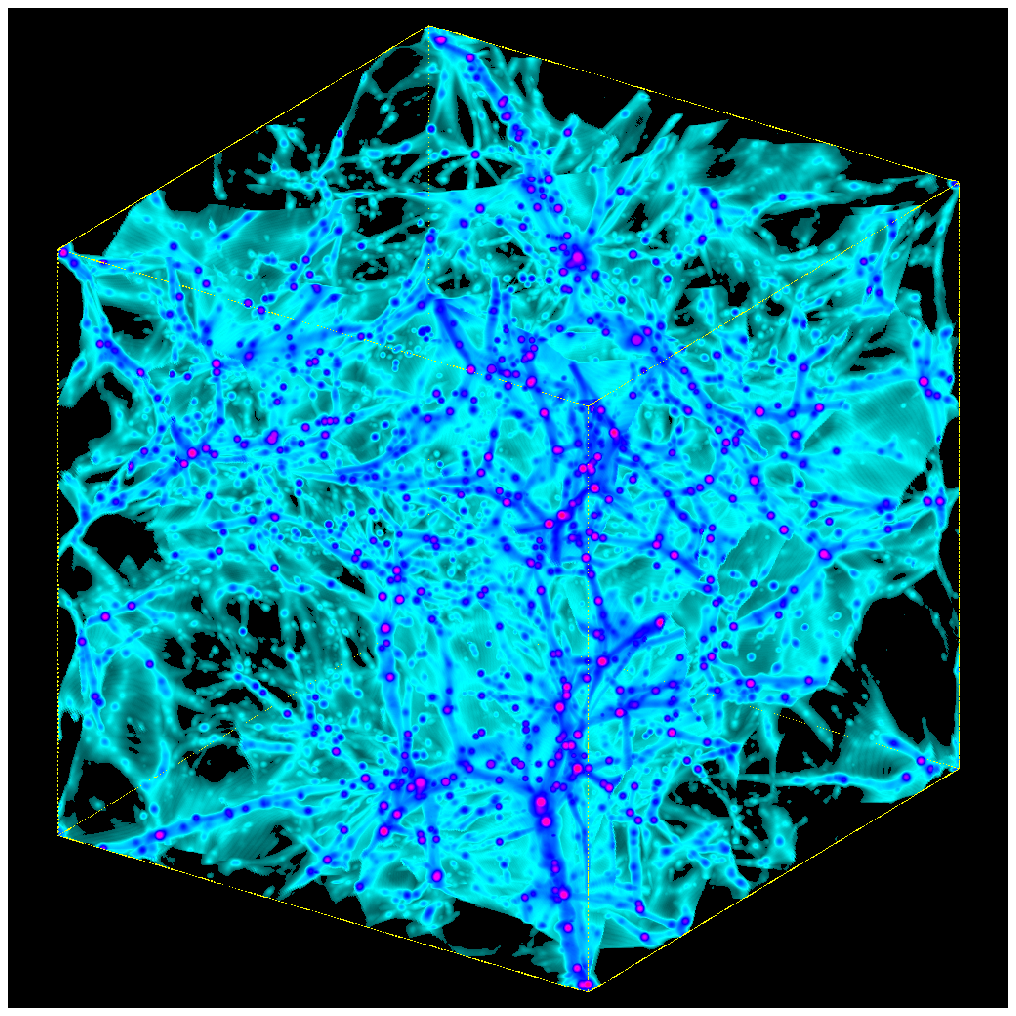}
\hspace{-3.0cm}
\includegraphics[width=0.68\textwidth]{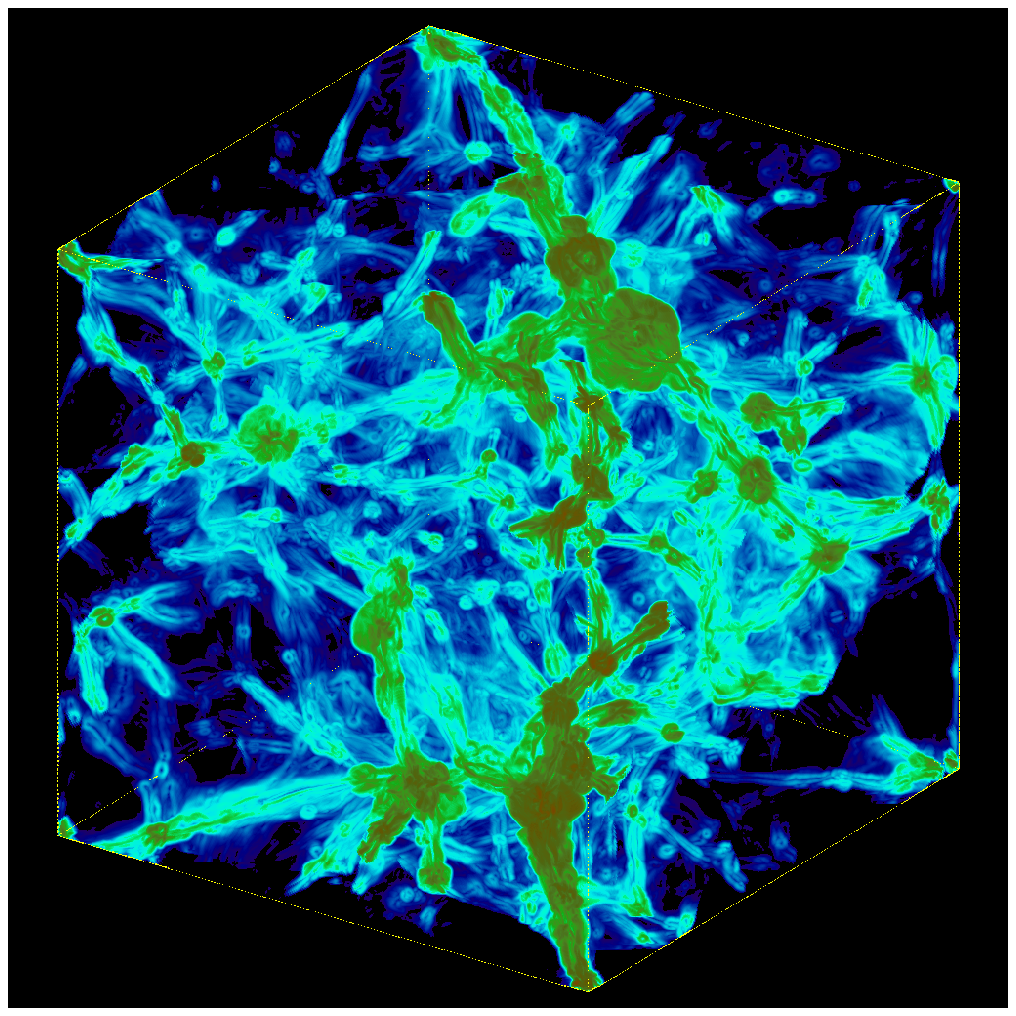}
\caption{Left: Projected view of the density of baryonic matter, $\rho_b$, at $z=0$ in simulation with box length of $25 \ h^{-1}$ Mpc; Right: The vorticity of velocity, $|\omega|$, in the same sample.}
\end{figure}

\subsection{Growth of Vortical Motions}
\begin{figure}[htbp]
\vspace{-1.0cm}
\hspace{-0.6cm}
\includegraphics[width=0.56\textwidth]{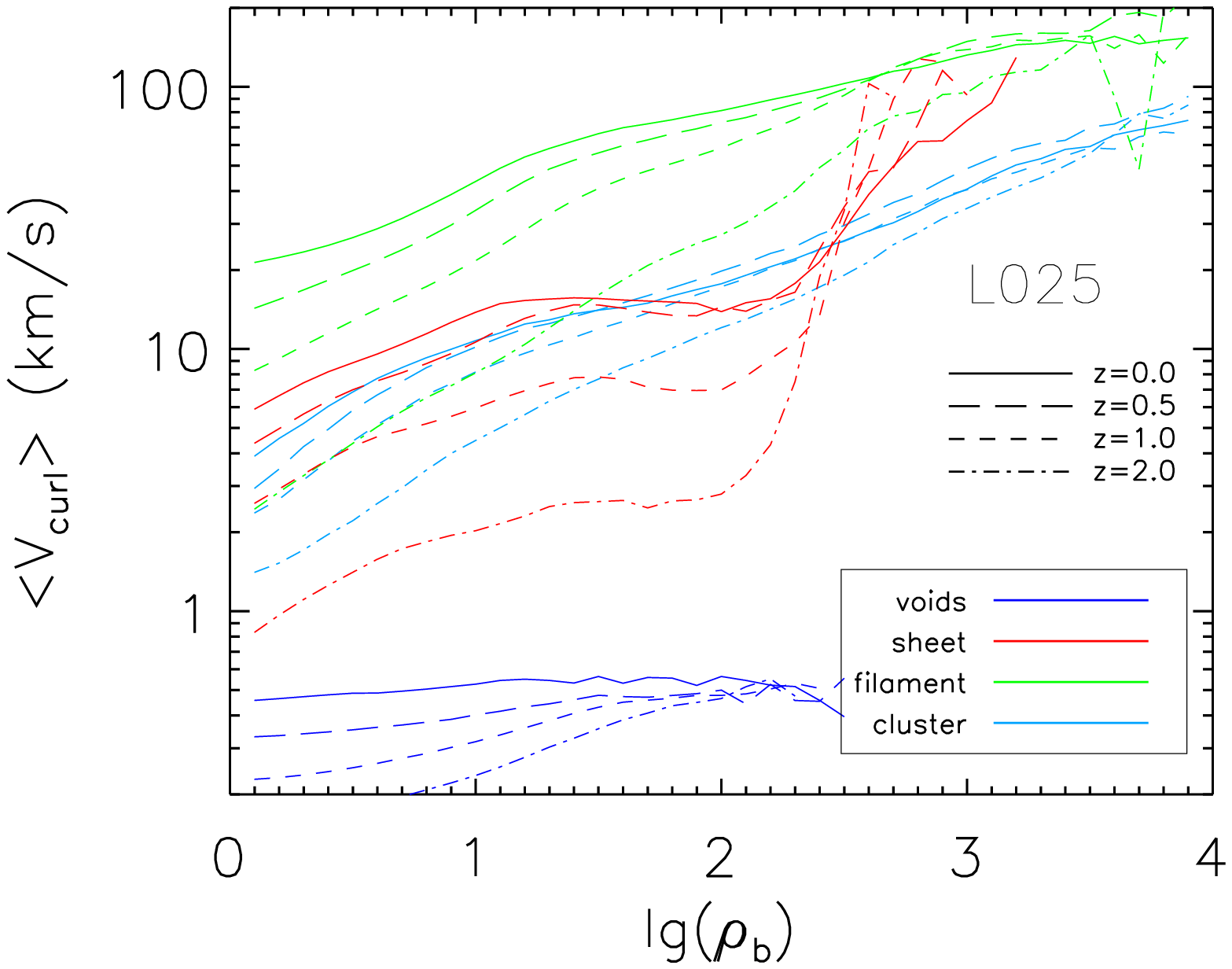}
\hspace{-1.60cm}
\includegraphics[width=0.56\textwidth]{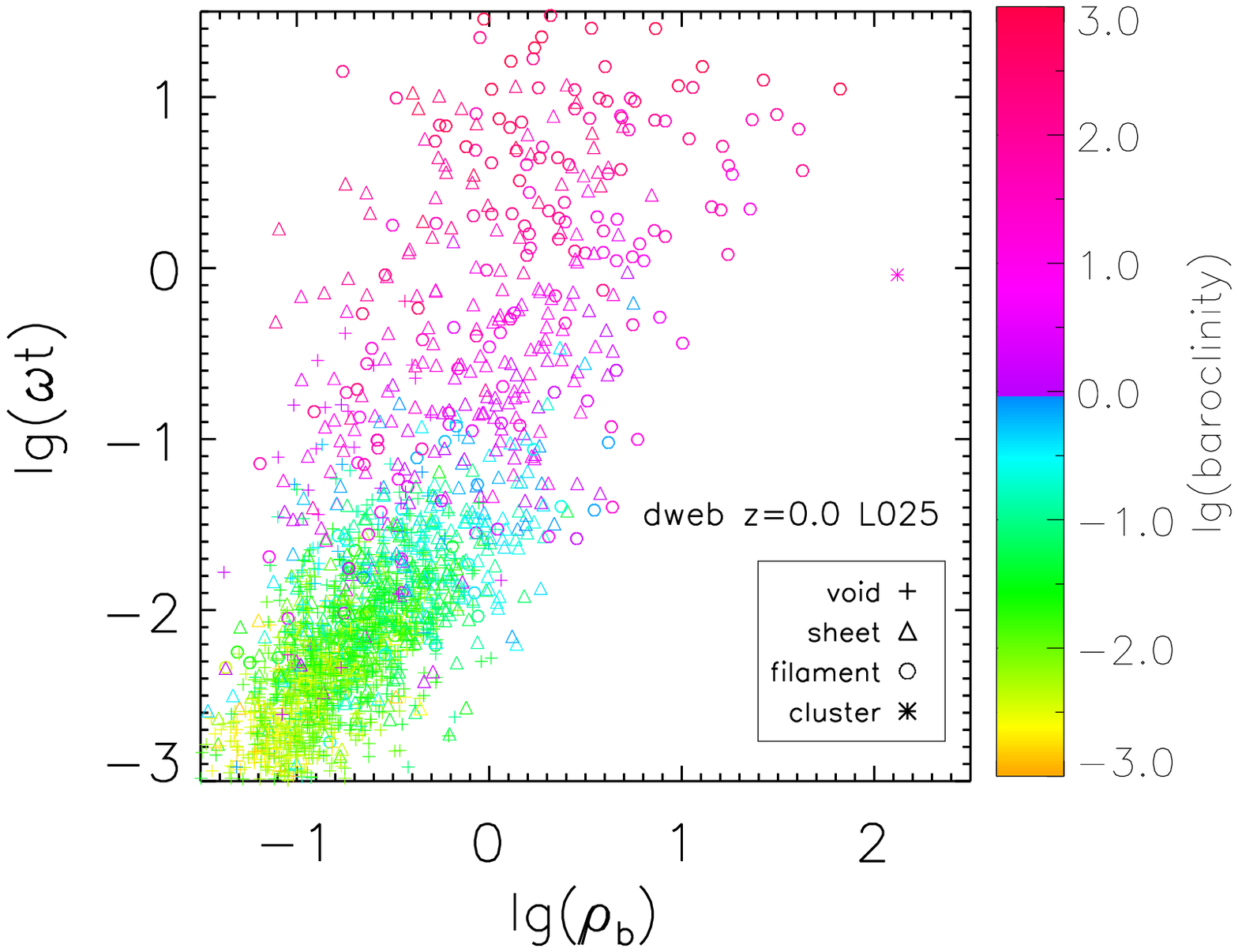}
\vspace{0.1cm}
\hspace{-0.6cm}
\includegraphics[width=0.56\textwidth]{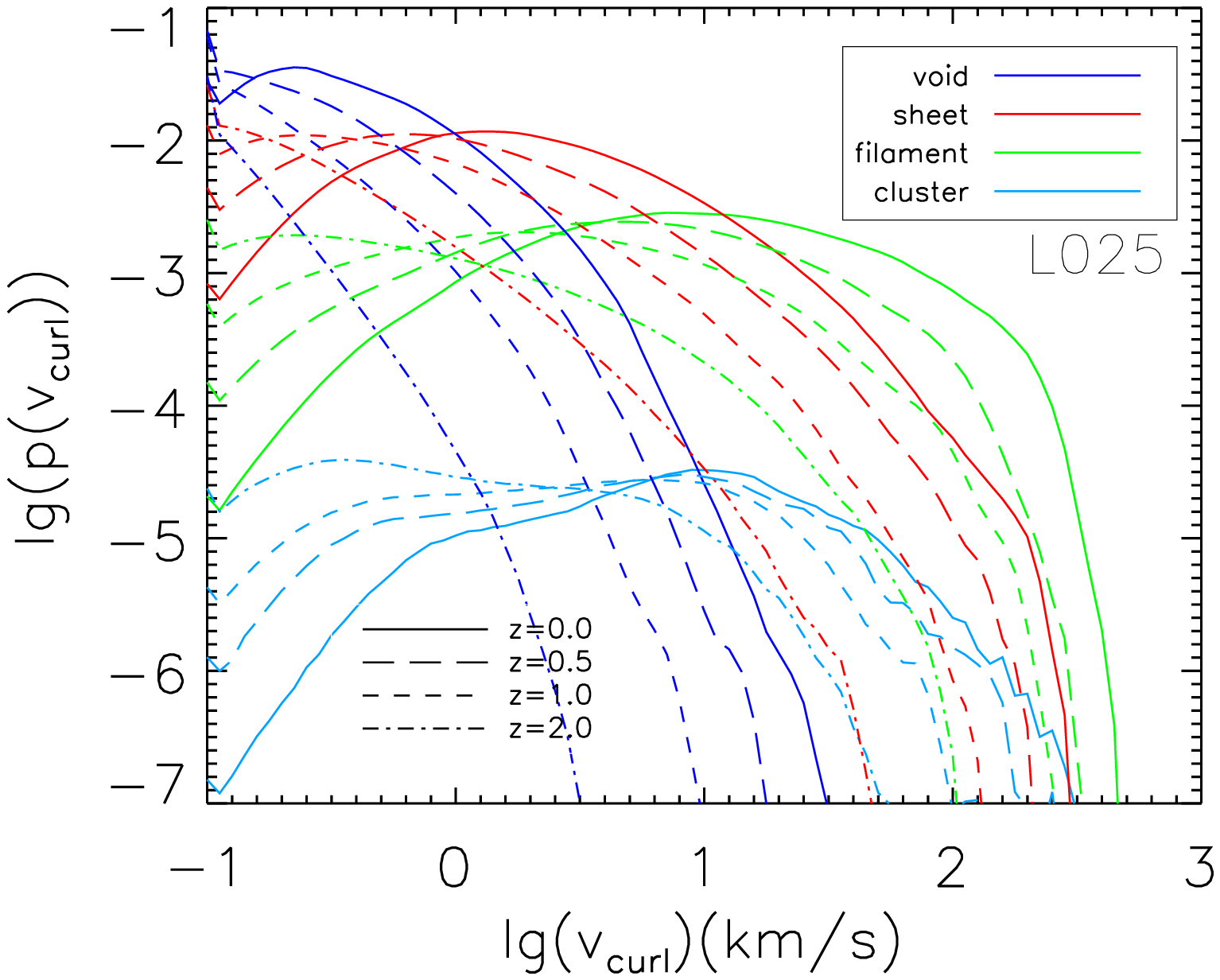}
\hspace{-1.6cm}
\includegraphics[width=0.56\textwidth]{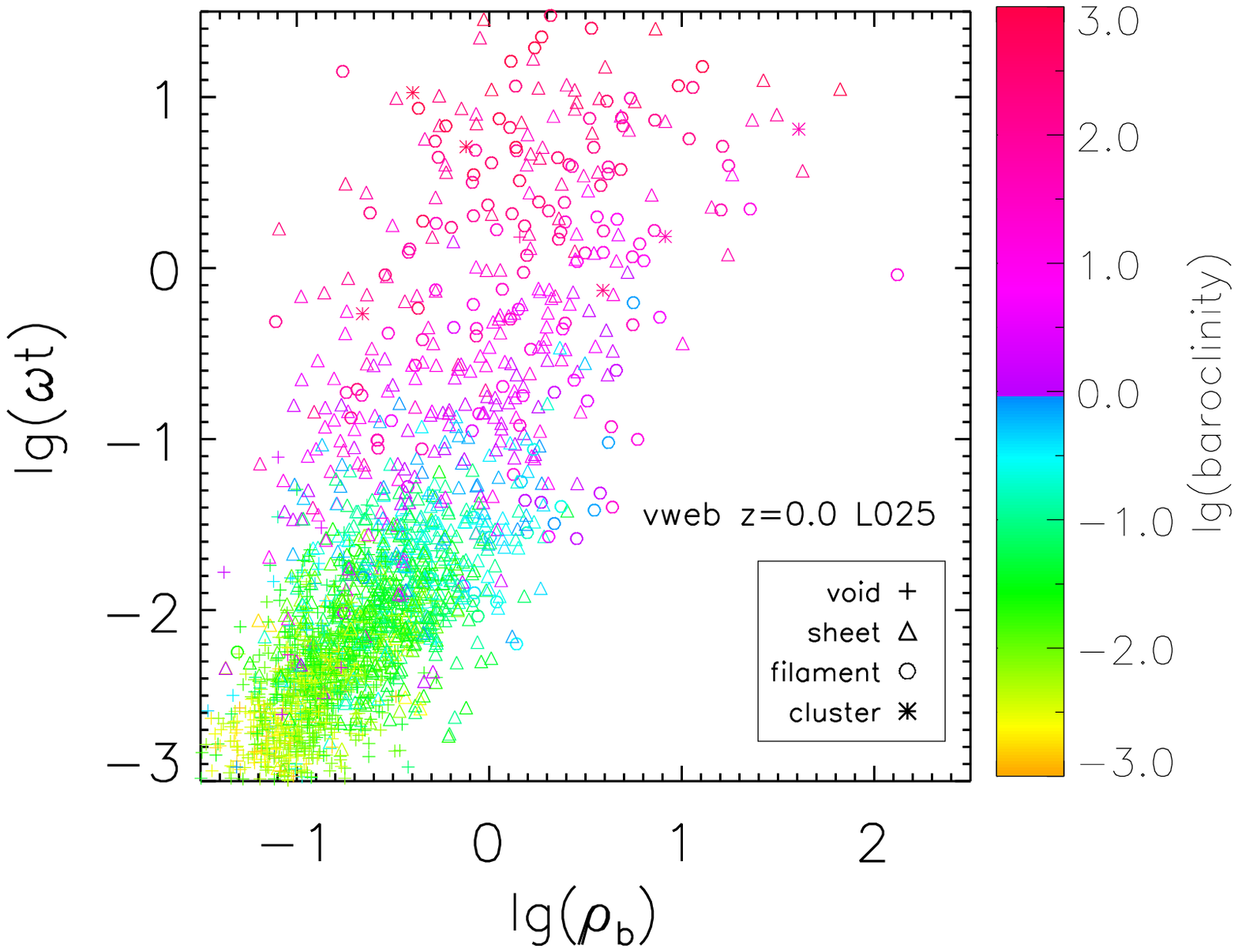}
\vspace{0.1cm}
\hspace{-0.6cm}
\includegraphics[width=0.56\textwidth]{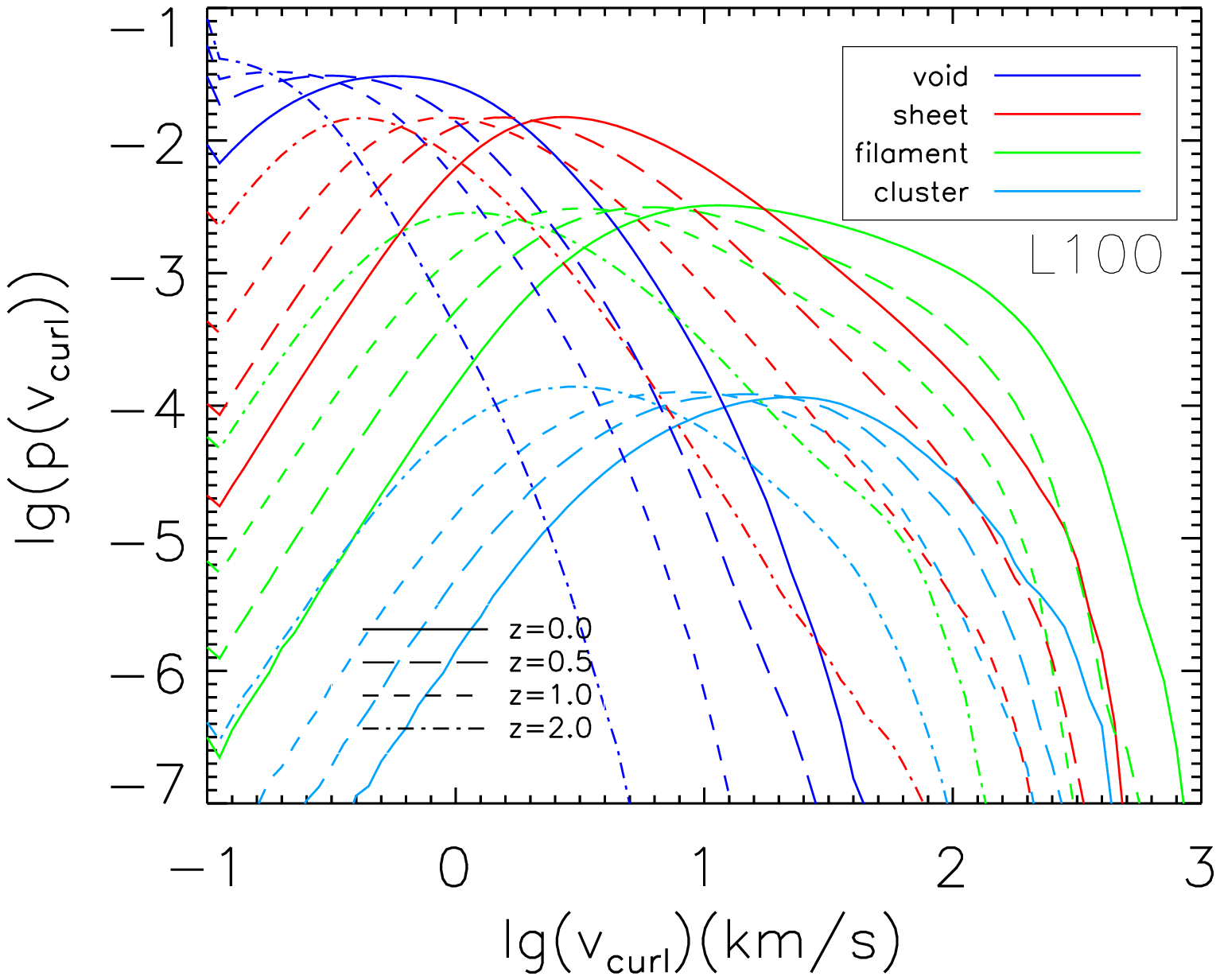}
\hspace{-1.6cm}
\includegraphics[width=0.56\textwidth]{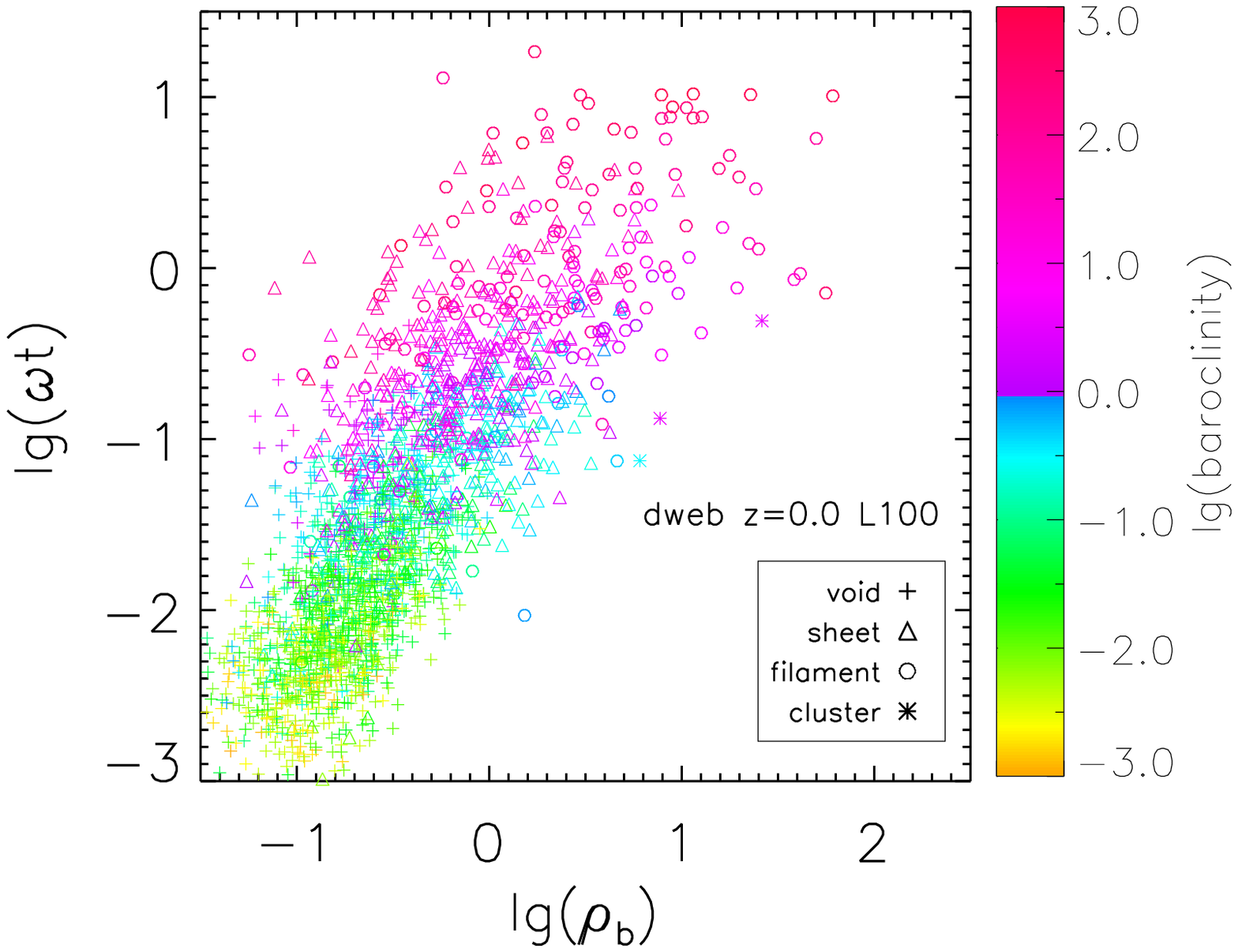}
\caption{Left column: evolution of density-weighted mean vortical velocity in L025(top); probability density function of vortical velocity in L025(middle); probability density function of vortical velocity in L100(bottom). Right column: the baroclinity of randomly selected grid cells in density-vorticity space with web classification using potential tidal tensor(\textbf{d-web}), and velocity shear tensor(\textbf{v-web}) in L025, and with (\textbf{d-web}) in L100, from top to bottom.}
\end{figure}

The left panel of Figure 1. presents a projected view of the density of baryonic matter, $\rho_b$,  in the simulation L025 at $z=0$, exhibiting clear visualization of voids, sheets, filaments and knots. The distribution of vorticity field, $|\omega|= |\nabla \times v|$, is demonstrated in the right panel of Figure 1. The vorticity has appears in sheets, i.e, the blue component in the right panel of Figure 1., which implies that vortical motions should be triggered as early as the gas flowing onto sheets. Comparing with the density map, the vorticity is developed significantly in the regions surrounding filaments and knots, and shows more extended shape. 

The divergence-free curl velocity, $\vec{v}_{curl}$,
associated with the vorticity can be separated from the total velocity through the Helmholtz-Hodge decomposition, $\vec{v}=\vec{v}_{curl}+\vec{v}_{div}$, 
where, $\nabla \cdot \vec{v}_{curl}=0$ and $\nabla \times \vec{v}_{curl} 
= \nabla \times \vec{v}$; $\vec{v}_{div}$ is the curl-free component, 
$\nabla \times \vec{v}_{div}=0$ and $\nabla \cdot \vec{v}_{div} = 
\nabla \cdot \vec{v}$. 
The evolution of mean curl velocity, $v_{curl}$, as a function of density in the cosmic web in L025 is shown in the top-left panel of Figure 2, in which the \textbf{d-web} identifier is used. The motions of baryonic matter in the voids are almost curl free and $v_{curl}$ is lower than $1 km/s$ even at $z=0$. The curl velocities are pumped up to $1-10 km/s$, and $10-150 km/s$ while the baryonic gas accreting onto sheets, and filaments. But when the gas flowing from filaments into knots, $v_{curl}$ is going to decrease, which should be related to dissipative processes during gravitational clustering. The evolution of the probability distribution function of $v_{curl}$ in the cosmic web since $z=2$, is illustrated in the middle-left panel of Figure 2. 
 
The governing equation of the modules of the vorticity, 
$\omega\equiv |{\vec \omega}|$, of baryonic matter reads as,
\begin{equation}
\frac{D \omega}{Dt}\equiv \partial_t {\omega} +\frac{1}{a}{\bf
v}\cdot{\nabla}\omega=\frac{1}{a}\left [\alpha \omega-d\omega +
\frac{1}{\rho^2}\vec{\xi}\cdot (\nabla \rho \times\nabla p )
-\dot{a}\omega \right ],
\end{equation} 
where $\vec{\xi}={\vec\omega}/\omega$, 
$\alpha={\vec\xi}\cdot({\vec\xi}\cdot\nabla){\bf v}$, and 
$d=\partial_iv_i$ is the divergence of the velocity field.
Vorticity is triggered by the 
baroclinity term, $(1/\rho^2)\nabla \rho \times\nabla p$, then stretched by the shear tensor, and in the meantime attenuated by the cosmic expansion (Zhu et al. 2010). 

In the right column of Figure 2, we present scatter plots in the density-vorticity space, where the scatter points stand for grid cells randomly selected from L025 at $z=0$, and are colored according to their values of baroclinity. In the up and middle panels,  the cosmic web are identified by \textbf{d-web} and \textbf{v-web} respectively. Evidently, there exists a positive correlation between the baroclinity and vorticity, in consistent with the deduction from equation (1). The growth history of vortical motions of gas depends on the evolution of baroclinity in different cosmic structures, which can be actually attributed to the emerging of curved cosmic shocks during gravitational collapse (Zhu et.al,  2010). 

The evolution of the probability distribution of $v_{curl}$, and the baroclinity in the density-vorticity space in L100 are also presented in the bottom panels of Figure 2. Obviously, the L025 and L100 simulations display a similar behavior on the overall evolution, though there appear some differences. Comparing with L025 simulation, the larger box size in L100 would produce more massive collapsed cosmic structures, and leads to the mild increase of the fraction of cells with large curl peculiar velocity in the bottom-left panel of Figure 2. On the other hand, the vorticity in L025 shows a slight upward shift, in contrast with L100 in the bottom-right panel of Figure 2. The higher resolution in L025 leads to less numerical dissipation and thus more sharply resolved cosmic shocks. In result, it brings on the enhancement of normalized vorticity in the post shock region, in consistent with the study of the impact of numerical viscosity in Zhu et al. (2013).

\section{The She-Leveque Scaling and Fractal Dimension}

The properties of velocity power spectrum of both the compressive and vortical motions below a couple of $Mpc$ were found to be consistent with the study on supersonic turbulence(Zhu et al. 2013). 
The more fundamental scaling properties concerned by theory of fully developed turbulence reads as
\begin{equation}
S_p(r) \ = \  < \delta v_r^p> = < |v(x)-v(x+r)|^p> \  \propto \ r^{\zeta_p},
\end{equation}
where $S_p(r)$ is known as the velocity structure functions. Kolmogorov(1941) proposed that there exist a universal scaling relation $\zeta_p=\frac{p}{3}$ in the inertial range of incompressible turbulence, i.e., the K41 law. However, a variety of laboratory experiments and numerical simulations have found that $\zeta_p$ deviates from the K41 law at large $p$, where intermittency corrections coming from the spatial fluctuations in the dissipation rate would be required . 

The intermittency model developed by She \& Leveque(1994), and then generalized by Dubrulle(1994) predicts a more universal scaling law for the relative scaling exponents $Z_p$ as a function of $p$, 
\begin{equation}
Z_p=\frac{\zeta_p}{\zeta_3}=(1-\Delta)\frac{p}{3}+\frac{\Delta}{1-\beta}(1-\beta^{p/3}),
\end{equation}
where the parameters $\beta$ and $\Delta$ depend on the hierarchy of intermittent structures during energy cascades. $C=\Delta/(1-\beta)$ is the co-dimension of the most intermittent structures and is related to the fractal Hausdorff dimension $d$ by $C=\mathfrak{D}-d$($\mathfrak{D}=3$ for three-dimension turbulence). The validity of the SL model have been verified in both incompressible and compressible turbulence, including supersonic turbulence(see, e.g. Kritsuk et al. 2007; Schmidt et al. 2008; Pan et al. 2009), turbulence in the star-forming interstellar medium(Boldyrev 2002; Federrath et al. 2010), as well as the cosmic baryon flow in the mildly nonlinear regime(He et al. 2006). 

\begin{figure}[tbp]
\vspace{-0.5cm}
\includegraphics[width=0.55\textwidth]{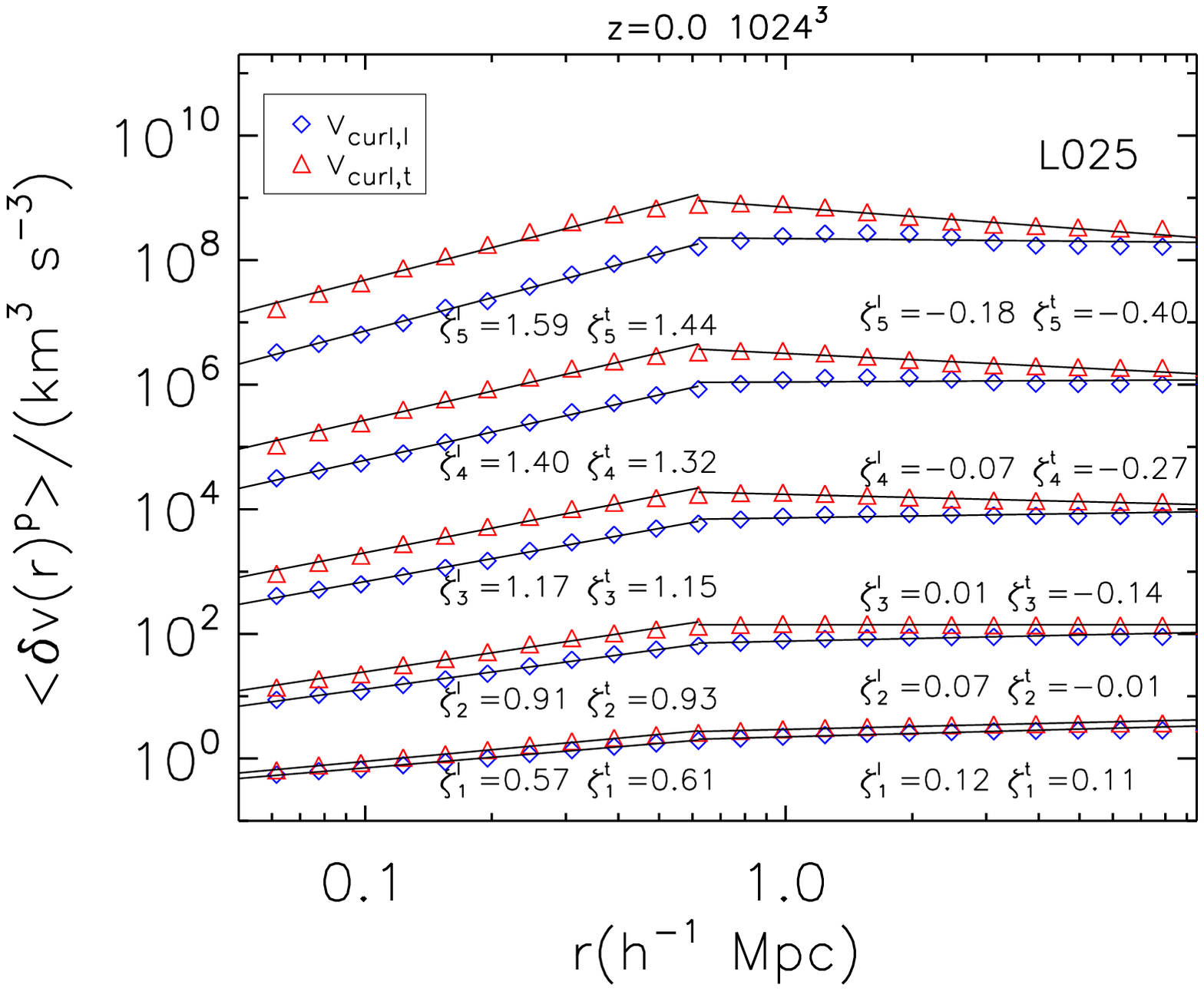}
\hspace{-0.5cm}
\includegraphics[width=0.55\textwidth]{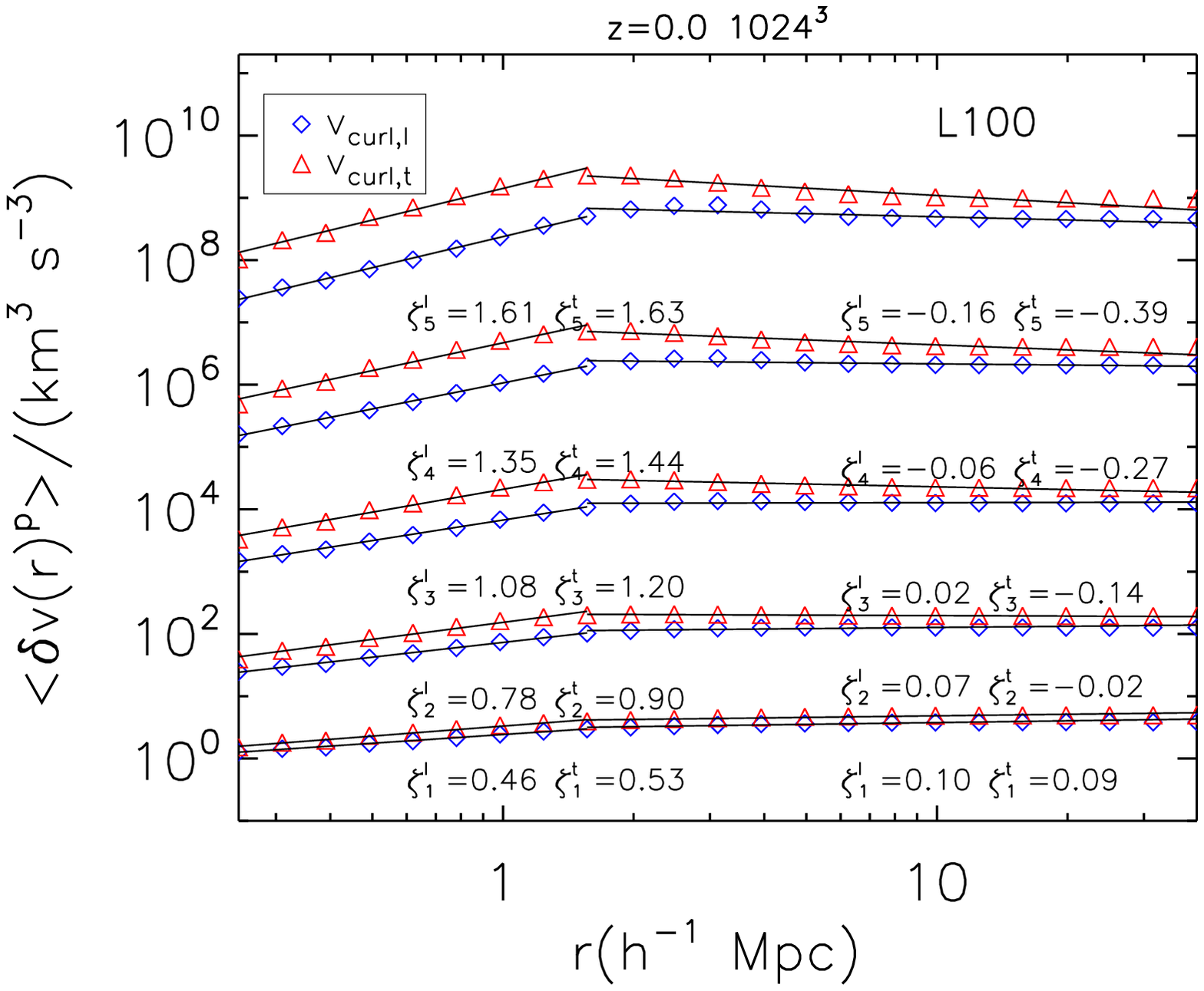}
\caption{Longitudinal and transverse velocity structure functions, $S_p^{l}$ and $S_p^{t}$ at $z=0$ in L025(left) and L100(right).}
\end{figure}

\begin{figure}[tbp]
\vspace{-0.5cm}
\includegraphics[width=0.55\textwidth]{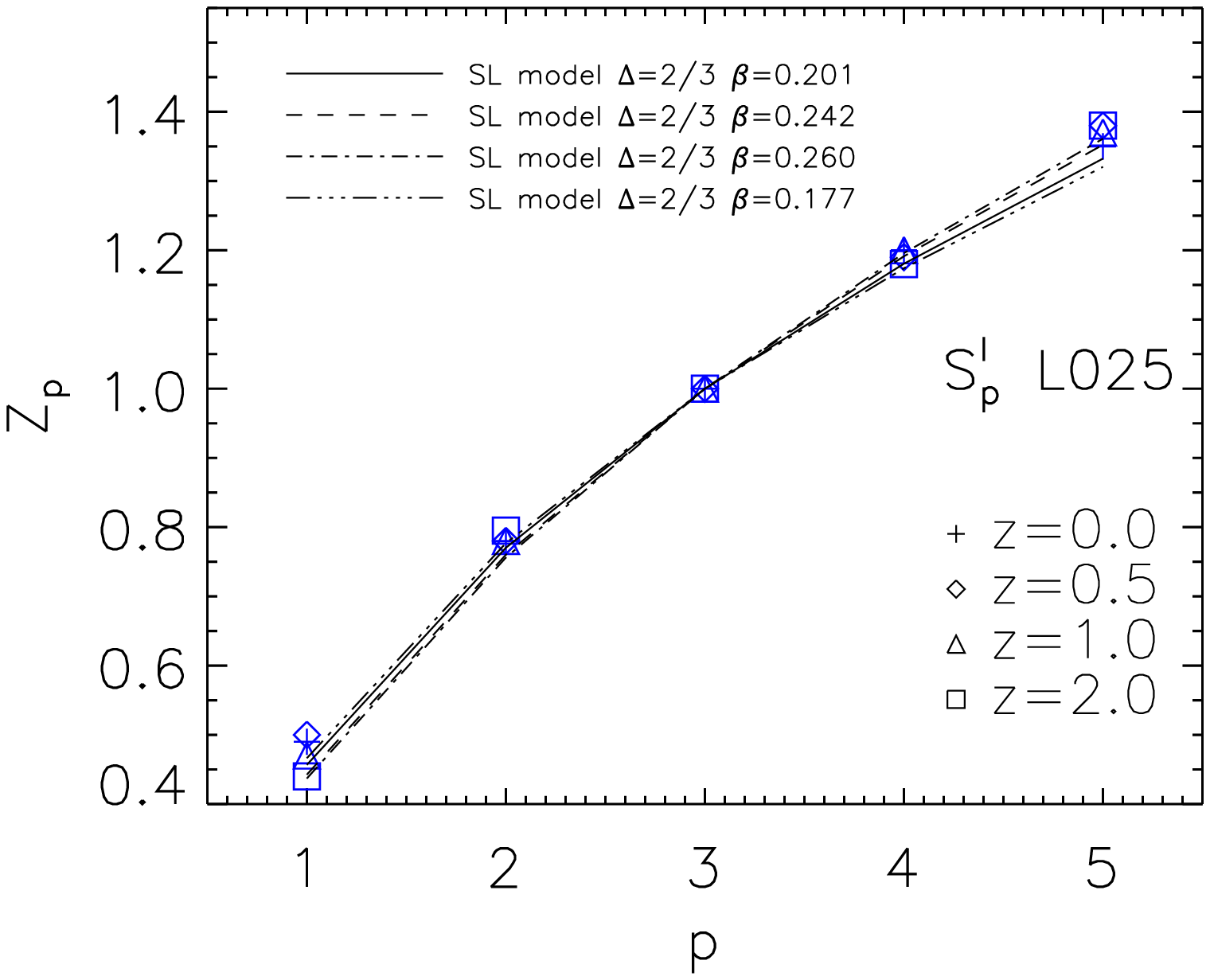}
\hspace{-0.5cm}
\includegraphics[width=0.55\textwidth]{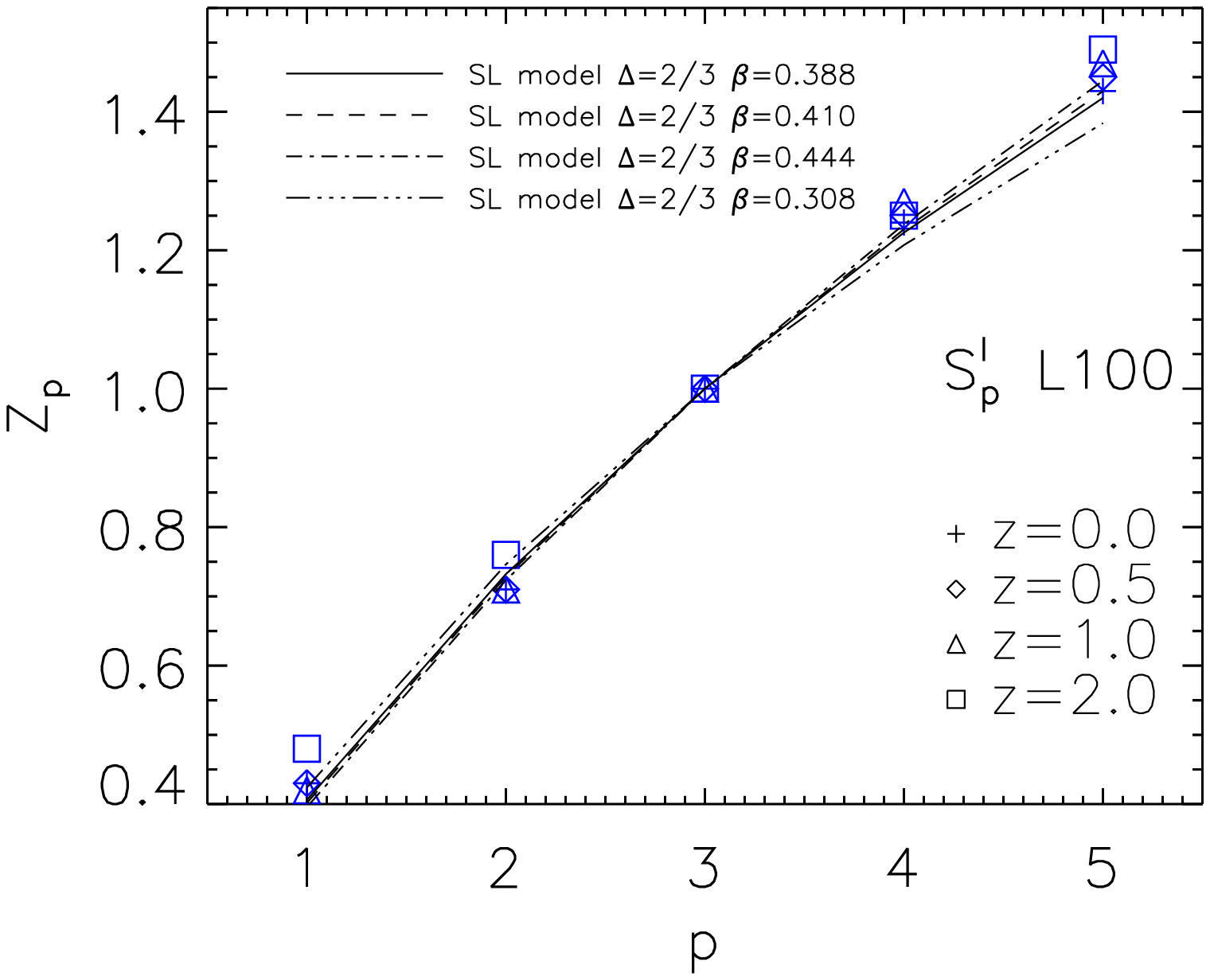}
\includegraphics[width=0.55\textwidth]{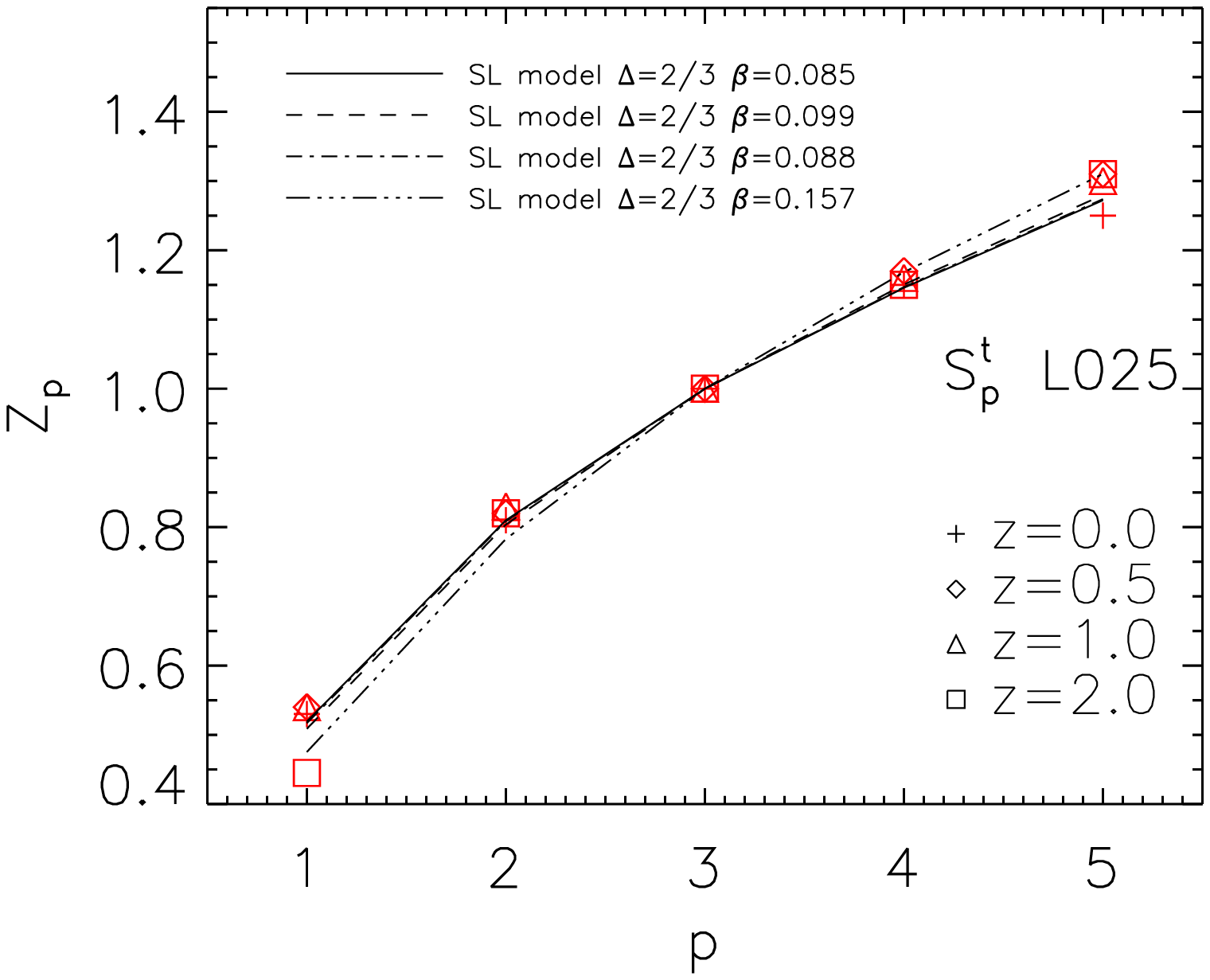}
\hspace{-0.5cm}
\includegraphics[width=0.55\textwidth]{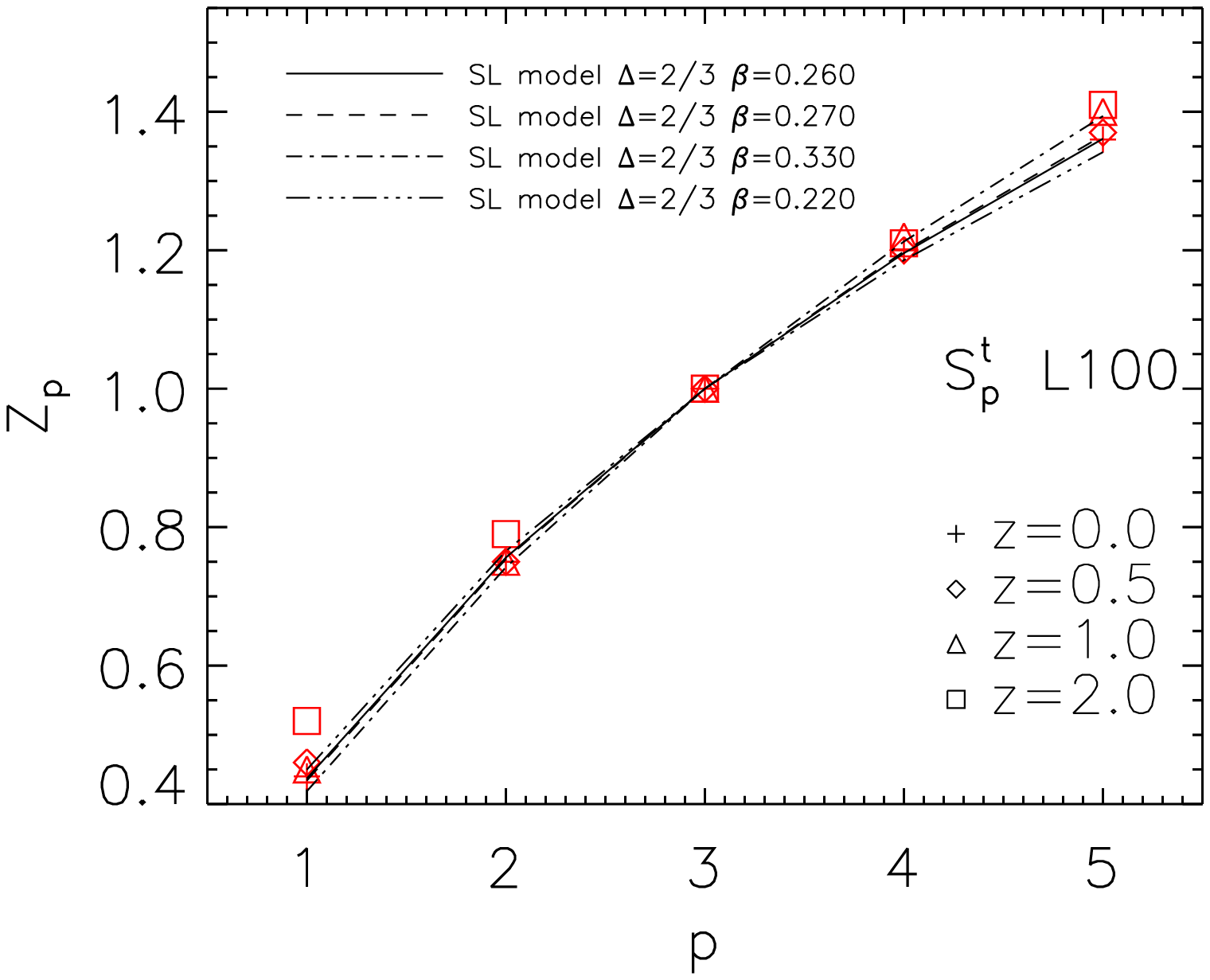}
\caption{Top(bottom) row: Relative scaling exponents $Z_p$ of longitudinal(transverse) mode since $z=2$, lines stand for the She-Leveque turbulence model with $\Delta=2/3$ and fitted $\beta$(Solid, dashed, dash-dotted, and triple-dot-dashed lines are fitting results at  z = 0, 0.5, 1.0, and 2.0). Left(right) column: result in L025(L100).}
\end{figure}

We analyze the velocity structure functions of curl velocity up to order $p=5$ since $z=2$, including the longitudinal ($\mathbf{v_{curl}} \| \mathbf{r}$) and transverse ($\mathbf{v_{curl}} \bot \mathbf{r}$) structure functions $S_p^{l}$ and $S_p^{t}$. The left panel of Figure 3 shows $S_p^{l}$ and $S_p^{t}$ at $z=0$ in L025. Obviously, there are two distinctive scaling regimes with a turning scale of $\sim 0.65 h^{-1}$Mpc, almost the same as the transition scale from subsonic to supersonic regime, i.e., $l\sim0.68 h^{-1}$ Mpc, inferred from the velocity power spectrum analysis (Zhu et al., 2013). We note that the turning scale in L100 is $\sim 1.5 h^{-1}$Mpc in Figure 3, right panel. Focusing on the supersonic regime, we plot the relative scaling exponents in both simulations in Figure 4. The simulation results are in good agreement with the predictions made by the SL model with $\Delta=2/3$, 0.08 $<\beta<$ 0.45. The Hausdorff dimensions given by the best-fitting are listed in Table 1, where the dimensions $d^l$ and $d^t$ correspond to $S_p^{l}$ and $S_p^{t}$ respectively. According to the SL model, the most singular structures of the vortical motions in the IGM have dimension of $2.10- 2.27$ and $1.80-2.10$ in L025 and L100 respectively. The regions with high vorticity are mainly composed of thick filaments intersecting at nodes as displayed in Figure 1, right panel.  Qualitatively, this visual result is in accordance with the deviation from the naive value of 1 (which would be applicable for very tiny filaments) in the dimensional analysis of those thick filaments, particularly in the scale range $l<0.65 h^{-1}$ Mpc. 

\begin{figure}[tbp]
\vspace{-0.5cm}
\hspace{-1.0cm}
\includegraphics[width=0.37\textwidth]{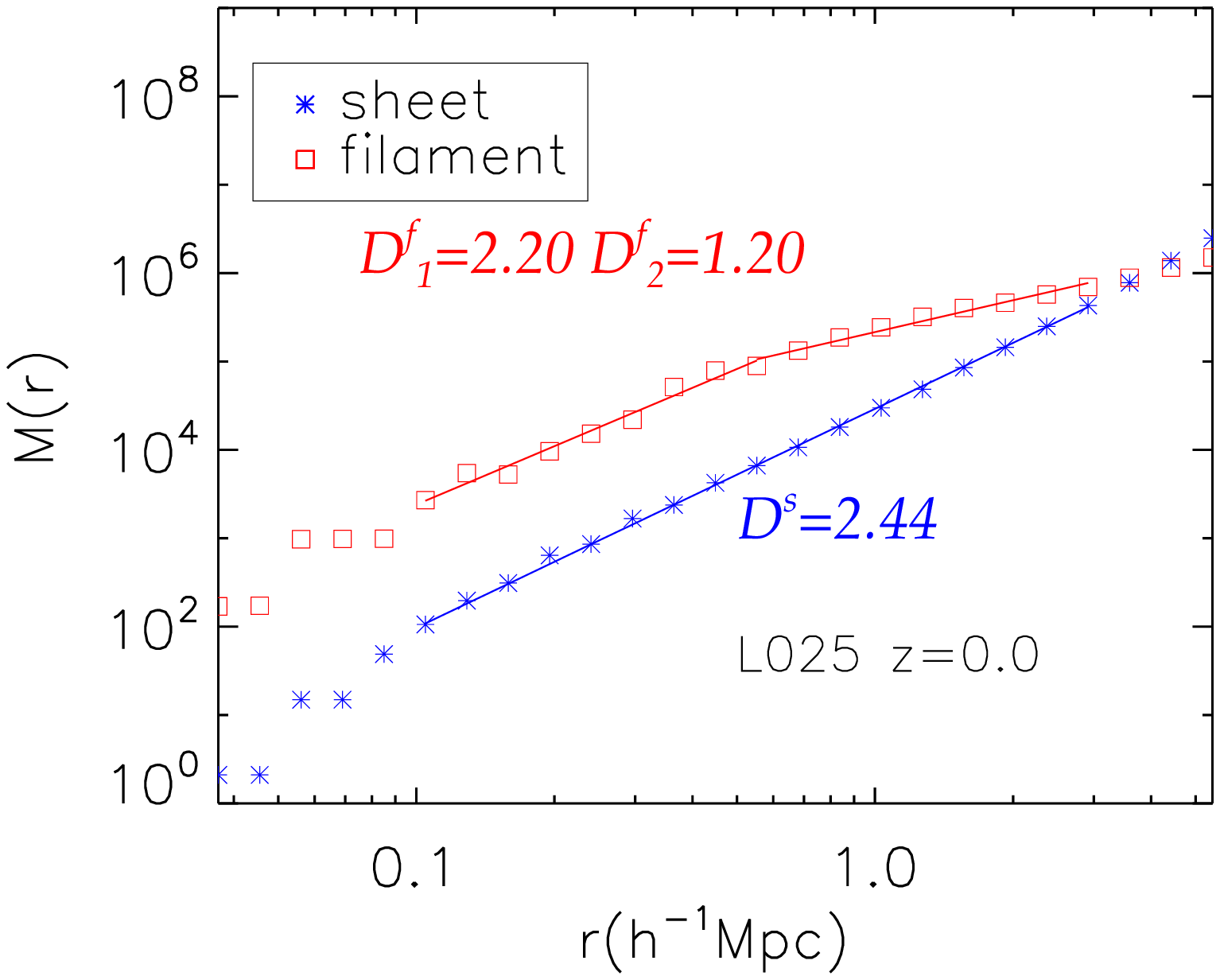}
\hspace{-0.8cm}
\includegraphics[width=0.37\textwidth]{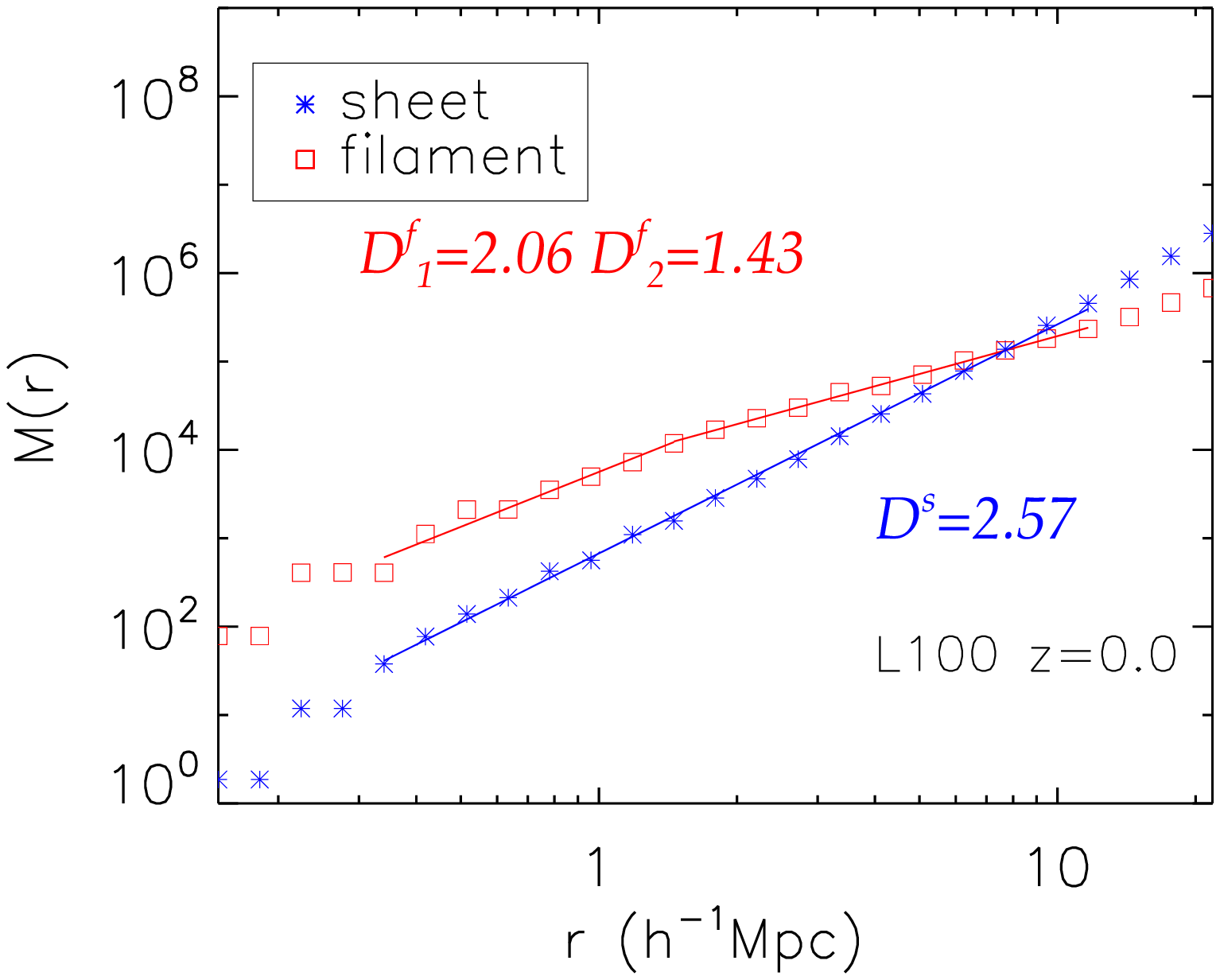}
\hspace{-0.6cm}
\includegraphics[width=0.37\textwidth]{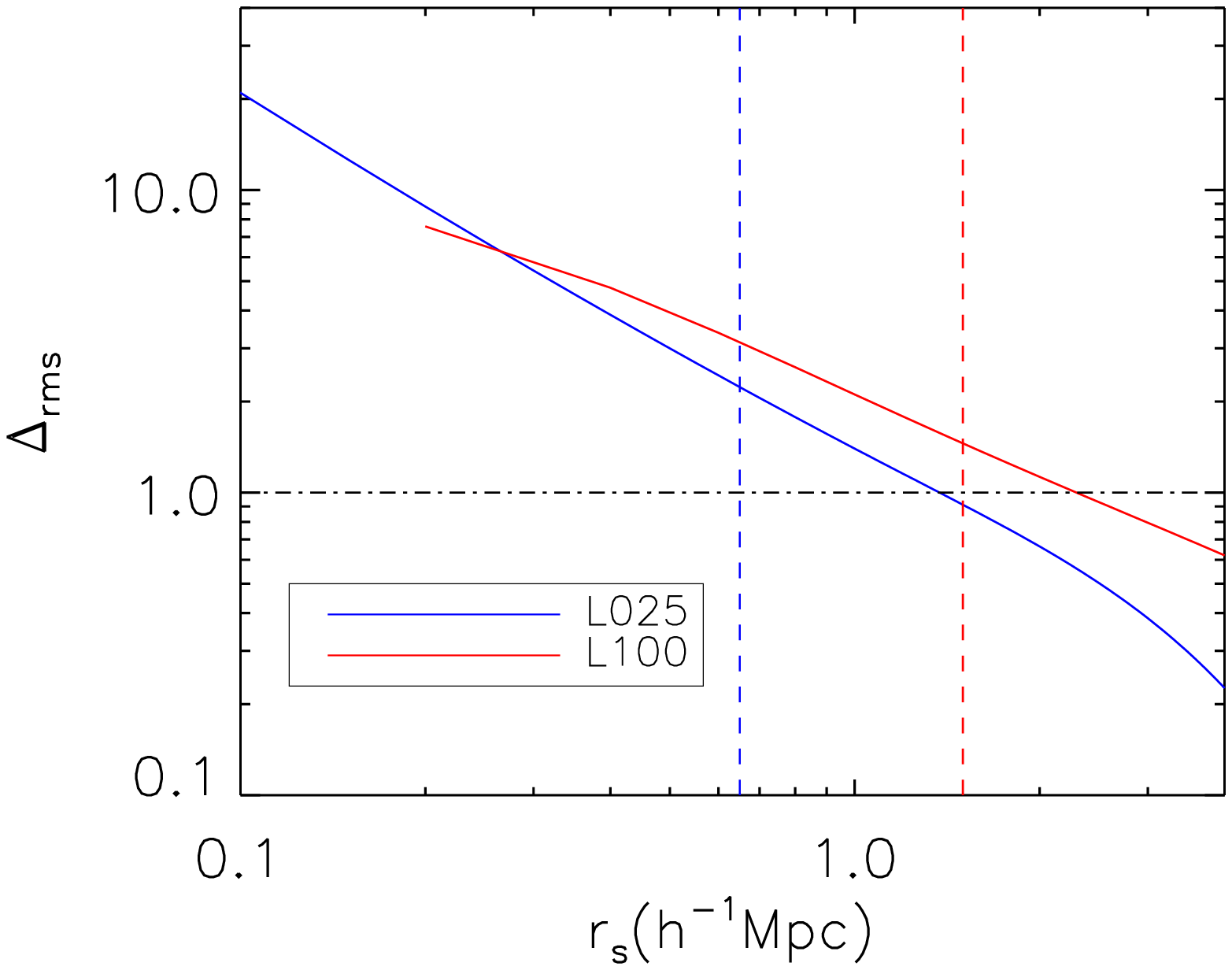}
\caption{Left(Middle): The fractal dimension of mass distribution in filaments and sheets at $z=0$ in L025(L100). Right: The rms of the density contrast field as function of smooth length. The vertical dash lines are plotted for the turning scales in both simulations. }
\end{figure}

The vorticity filaments are highly correlated with those filaments in the matter distribution. The fractal dimensions of the mass distribution in the cosmic web can be estimated directly by the box counting method (Mandelbrot 1983), which has been applied to studies of supersonic turbulence (e.g., Kritsuk et al. 2007), and the cosmic web (Cautun et al. 2014). We follow the way of Kritsuk et al.(2007) to compute the mass dimensions of filaments and sheets. Grid cells with $\rho_b$ larger than $90\%$ of those grid cells in the filaments(sheets) identified by the \textbf{d-web} are chosen as the center of boxes of length $r$. We sum the gas mass in those grid cells which are within boxes of length $r$ and assigned to filaments (sheets) to counting the averaged mass $M(r)$, in the estimation of $M(r) \propto r^{\textit{D}}$ of the filaments (sheets). 

\begin{table}[htbp]
\begin{center}
\begin{tabular}{|c|c|c|c|c|c|c|c|c|}
\hline
\multirow{2}{*}{Redshift} & \multicolumn{4}{|c|}{L025} & \multicolumn{4}{|c|}{L100}\\
\cline{2-9}
& $d^{l}$ & $d^{t}$  & $\textit{D}^{f}$ & $\textit{D}^{s}$ &  $d^{l}$ & $d^{t}$  & $\textit{D}^{f}$ & $\textit{D}^{s}$ \\
\hline
0.0 & 2.17 & 2.27  & 2.20 & 2.44  & 1.91 & 2.10  & 2.06 & 2.57\\
\hline
0.5 & 2.12 & 2.26  &  2.06 & 2.49 &  1.87 & 2.10 & 1.96 & 2.62 \\
\hline
1.0 & 2.10 & 2.27 & 1.80 & 2.55 &  1.80 & 2.01 &  1.90 & 2.67\\
\hline
2.0 & 2.19 & 2.20  & 1.85 & 2.61  & 2.03 & 2.13 &  1.94 & 2.70\\
\hline
\end{tabular}
\caption{The Hausdorff dimension $d$ derived from fitting of relative scaling exponents in the She-Leveque model ($d^{l}$ for longitudinal and $d^{t}$ for transverse components of velocity fields), and fractal dimension $\textit{D}$ of mass distribution in sheets ($\textit{D}^s$) and filaments($\textit{D}^f$) since $z=2$.}
\end{center}
\vspace{-0.5cm}
\end{table}

{The left panel of Figure 5.} shows $M(r)$ of filaments and sheets in L025 at $z=0$. We focus our analysis in the scale range of $0.1-3.0 Mpc$, in which the velocity field shows behaviors of fully developed turbulence. Obviously, $M(r)$ of filaments also displays a transition from $\textit{D}^f=2.20$ to $\textit{D}^f=1.20$ at the turning scale $\sim 0.65 h^{-1}$Mpc, which is very close to $S_p(r)$. A similar transition can be found in L100 at $\sim 1.50 h^{-1}$Mpc, with the dimensions listed in Table 1. While in sheets, the transition is too weak to observe, and the fractal dimension $\textit{D}^s$ keeps almost a constant in the studied scale range. The turning scale appears to be dependent on the characteristic cross-section radius (thickness) of filaments (sheets). Once the box length $r$ exceeds the cross-section radius of filaments and keeps increasing, those filaments are closing to idealized thin filaments with dimension $\textit{D}^f=1$. To see the effect of simulation box size, we compare the rms  $\Delta_{rms}$ of density contrast  $(\rho-\bar{\rho})/\bar{\rho}$, as a function of smooth length in both simulations in the right panel of Figure 5. At the turning scales,  $\Delta_{rms} \sim 2.1$ in L025 and $\Delta_{rms} \sim 1.6$ in L100. The lack of larger scale perturbations in L025 may result in a narrower nonlinear regime, and probably leads to thinner filaments. Since the vorticity are mainly produced through curved shocks surrounding filaments and knots during gravitational collapse, it is easy to understand the similarity between the turning scales extracted from the vorticity and density fields. 

The variation of the fractal dimensions of mass distribution along with redshifts is also shown in Table 1. The fractal dimension $\textit{D}^s$ of gaseous sheets in our simulation is slightly larger than the result inferred from the dark matter simulation (Cautun et al. 2014), i.e. $\sim 2.40$. The fractal dimension $\textit{D}^f$ of filaments, however, is slightly smaller than Cautun et al.(2014). These minor discrepancies might be due to cosmic variances, the difference between estimation methods, and the distributions of baryon and dark matter. The Hausdorff dimensions derived from the velocity structure functions are larger than the fractal dimensions of density filaments by $\sim 0.05-0.3$ in most cases, except that from $S_p^l$ in L100. Visually, the vorticity filaments are more extended than the density filaments. 

It is reasonable to believe that the turning scale inferred from the L100 simulation would be more close to the real value in cosmic matter distribution. It is simply because the larger simulation box allows the growth of longer wavelength perturbations and leads to formation of more massive collapsed cosmic structures, and hence larger characteristic size of structures and associated supersonic regime. On the other hand, the Hausdorff dimension derived from L025 would be more reliable, since Hausdorff dimension depends on the asymptotic behavior of mass and velocity distribution on small scales, the higher space/mass resolution will be helpful to resolve small scales structures including curved shocks, the local density and velocity fields, more precisely.

\section{Discussion and Conclusion}

The growth of vortical motions of baryonic matter in the cosmic web, and the scaling properties of curl velocity fluctuations have been studied using cosmological hydrodynamical simulations in this work. We find that,

1. The vortical motions are triggered by the baroclinity around shocks when baryonic gas flowing from void onto sheets, and especially from void and sheets onto filaments, and then experience dissipation while entering into the knots. The mean curl velocities are about $< 1$, $1-10$, $10-150$, $5-50$ kms$^{-1}$ in voids, sheets, filaments and knots at $z=0$, respectively. 

2. The vortical motions of the IGM can be well described by the She-Leveque(SL) turbulence model via the velocity structure functions up to order of 5 in the scale range of $l<0.65(1.50) h^{-1}$ Mpc in the simulation of box size 25(100) $h^{-1}$ Mpc. The fractal dimensions of vorticity of the IGM that derived from the SL model with $\Delta=2/3$ are $d \sim 2.1-2.3(1.8-2.1)$ at $z<2$.

3. In the filaments, there are two distinctive mass scaling regimes with the turning scale of $0.65(1.50)$ $h^{-1}$ Mpc in the simulation of 25(100) $h^{-1}$ Mpc, consistent with those inferred from the She-Leveque scaling of curl velocity fluctuations. Physically, the turning scale characterizes the transition from the subsonic to supersonic regime, and is likely related to the typical radius of filaments, around which, numerous curved shocks are formed. The fractal dimension of mass distribution in the filaments are $\textit{D}^f \sim 1.9-2.2$ in the supersonic regime, and is smaller than $d$ by $\sim 0.05-0.3$ in most cases. While for sheet-like structures, the transition is very weak over the full relevant scales, and $\textit{D}^s \sim 2.4-2.7$. This result does imply that the dominant structures in the process of vortical kinetic energy transportation during gravitational clustering would be thick vorticity filaments. 

The IGM play an important role in the formation of galaxies, and hence in shaping their properties. Recent simulation works suggest that spins of halo and galaxy show alignment with large scale vorticity of dark matter and gas respectively (Libeskind et al 2013; Dubois et al. 2014; Laigle et al. 2015). Our work provides an alternative view on the development of vorticity when baryonic gas flow into the cosmic web. Limited by resolutions, this work is unable to trace the dynamics of vorticity on the scale of galaxies, and make a connection to galaxies spins. A comparison study between dark matter and baryonic gas on the development and distribution of vorticity would be helpful to understand the alignment between vorticity and large scale structures, which is feasible and will be reported in the next work. 

The hierarchy of vorticity filaments is associated with the formation of density filaments, i.e., gas in voids and sheets with low vorticity flowing onto filamentary structures. This scenario allows a simple physical interpretation of the SL model. The fractal dimensions of vorticity in the supersonic regime of the IGM, $d \sim 2.1-2.3$, are quite close to the values reported by several supersonic turbulence studies, e.g., in interstellar gas by simulations and observations (Boldyrev 2002;  Federrath et al. 2010). In their works, the shock fronts and hence sheets emerging in gravitational collapse are identified as the primary structures. However, our study indicates that the dominant structures in the IGM should be thick filaments. We have taken a fixed value of $\Delta=2/3$ in the SL model. More comprehensive study on the values of $\Delta$ and $\beta$ is related to the problem known as the universal scaling relation in supersonic turbulences (Schmidt et al. 2008; Pan et al. 2009; Federrath et al. 2013). 

Finally, we need to point out that both the scale range of turbulence and the fractal dimensions of vorticity and mass distribution show some differences in our simulations. The length and diameter of filaments, the thickness of sheets, and their evolution may largely contribute to these differences. Recent investigations by other groups also show that the properties of cosmic filaments and sheets are dependent on the box sizes, resolutions and identification methods (Colberg et al. 2005; Choi et al. 2010; Arago-Calvo et al. 2010; Cautun et al. 2014). To resolve this discrepancy, large simulations with high spatial resolution,  e.g., hundreds of Mpc in box size and tens of kpc in resolution, would be necessary. The feedback from star formation, and AGN, which are not followed in this work, are also needed in order to provide a precise description of the vortical motions of baryonic gas down to galactic scales.

\begin{acknowledgements}
Acknowledgements: 
\end{acknowledgements}
The authors thank the anonymous referee for very useful comments. The simulations were run at Supercomputing Center of the Chinese Academy of Sciences, and SYSU. WSZ is supported under the National Natural Science Foundation of China(NSFC) grant 11203012. FLL is supported under the NSFC grants 11273060, 91230115 and 11333008, and State Key Development Program for Basic Research of China (No. 2013CB834900 and 2015CB857000).

\end{document}